\documentclass[10pt]{article}
\usepackage[a4paper,top=30mm,bottom=30mm,left=20mm,right=20mm]{geometry}
\usepackage{graphicx}
\usepackage{amsmath}
\usepackage{amssymb}
\usepackage{booktabs}
\usepackage{multirow}
\usepackage{cite}
\usepackage{url}
\usepackage{newtxtext}
\usepackage{newtxmath}
\usepackage{caption}
\usepackage{framed}
\usepackage{fancyvrb}
\usepackage[most]{tcolorbox}  
\usepackage{fvextra}          
\usepackage{enumitem}         

\tcbuselibrary{skins, breakable}

\DefineVerbatimEnvironment{AlgorithmCode}{Verbatim}{
  fontsize=\small,
  breaklines=true,
  breakanywhere=true,
  breaksymbol={},
  numbers=left,           
  numbersep=6pt,          
  xleftmargin=1em         
}

\tcbset{
  algobox/.style={
    enhanced,
    breakable,
    colback=white,
    colframe=black,
    boxrule=0.4pt,
    toprule=0.4pt,
    bottomrule=0.4pt,
    leftrule=0.4pt,
    rightrule=0.4pt,
    arc=0pt,
    outer arc=0pt,
    left=2mm,
    right=2mm,
    top=2mm,
    bottom=2mm,
    toprule at break=0.4pt,
    bottomrule at break=0.4pt,
    pad at break=2mm,
  }
}

\title{\LARGE\bfseries AFLL: Real-time Load Stabilization for MMO Game Servers \\
Based on Circular Causality Learning}

\author{
\large Shinsuk Kang$^{\mathrm{O}}$, Youngjae Kim$^{\dagger}$ \\[2mm]
\normalsize Graduate School of AI and Software, Sogang University \\[1mm]
\normalsize \texttt{kang.shinsuk@gmail.com, youkim@sogang.ac.kr} \\[2mm]
\footnotesize $\dagger$ Corresponding Author
}
\date{}

\begin{document}

\maketitle

\begin{abstract}
Massively Multiplayer Online (MMO) game servers experience performance degradation due to high message processing loads generated by thousands of concurrent users. In MMO games, state synchronization between players is essential, and the circular causality where server-transmitted messages trigger additional player requests forms feedback loops that rapidly amplify the load. While existing research has focused on network bandwidth reduction and processing efficiency, no prior work has learned and controlled feedback loops during runtime. This paper proposes the Adaptive Feedback Loop Learning (AFLL) system utilizing backpropagation. AFLL tracks the impact of each message type on the load in real-time and dynamically adjusts transmission rates using gradient descent to prevent load spikes proactively. Experimental results with 1,000 concurrent users show that AFLL achieves 48.3\% reduction in CPU time (6,143ms $\rightarrow$ 3,177ms), 64.4\% reduction in thread contention (60.36\% $\rightarrow$ 21.51\%), and substantial reduction in performance spikes (99.7\% reduction in spikes exceeding 50ms) compared to learning disabled. Both Learning OFF (baseline) and Learning ON demonstrated excellent reproducibility with coefficients of variation of 1.83\% and 0.54\%, and all three Learning ON experiments converged to identical weights.
\end{abstract}

\vspace{3mm}

\section{Introduction}

Massively Multiplayer Online (MMO) game servers have the characteristic that message transmission load is rapidly amplified due to state synchronization among thousands of concurrent users. Throughout this paper, 'player' refers to a game user, while 'client' refers to the game client program. The circular causality, where server-transmitted messages trigger additional player requests, forms feedback loops that rapidly amplify the load, resulting in server performance degradation, response delay spikes, game fairness collapse, and player churn. This problem threatens the sustainability of game services beyond causing technical performance issues. While existing research has focused on network bandwidth reduction and processing efficiency, no prior work has learned and controlled feedback loops during runtime.

The goal of this research is to implement a real-time adaptive system that learns circular causality to prevent load spikes proactively. To this end, this paper proposes the Adaptive Feedback Loop Learning (AFLL) system utilizing backpropagation. AFLL learns the load contribution of each message type in real-time and achieves proactive control with zero learning overhead by parallelizing learning and control. In experiments with 1,000 concurrent users, it achieved 48.3\% CPU reduction (6,143ms $\rightarrow$ 3,177ms), 64.4\% thread contention reduction (60.36\% $\rightarrow$ 21.51\%), and substantial spike reduction (99.7\% reduction in spikes exceeding 50ms), demonstrating excellent reproducibility for both Learning OFF and Learning ON (coefficients of variation of 1.83\% and 0.54\%).

\section{Background and Research Motivation}

\subsection{Characteristics of MMO Game Servers}

MMO (Massively Multiplayer Online) game servers are high-load real-time distributed systems where numerous users simultaneously connect and interact. In MMO games, to maintain consistent game state, each player's position, actions, and state changes must be synchronized in real-time to all nearby players. Synchronization delays prevent proper gameplay. For example, if player A attacks an enemy but synchronization failure due to server load makes the attack appear to miss to other players, game fairness is compromised, leading to player churn\cite{mmo1}. As the number of players $(n)$ increases, the amount of state information to transmit grows as $O(n^2)$, specifically $n \cdot (n-1)$. Each player must transmit their state to all other nearby players, so in a region with 100 concentrated players, $100 \times (100-1) = 9,900$ state transmissions are required. When hundreds gather in a specific area during large-scale battles or events, instantaneous load spikes occur, and as the server approaches processing limits, response delays surge, connections drop, and gameplay experience deteriorates.

\subsection{The Circular Causality Problem}

The core of this problem is not simply the large traffic volume, but rather the \textbf{circular causality} where server-transmitted messages trigger additional player requests. This differs fundamentally from the linear load scaling in typical web servers. For example, when the server transmits information about 10 nearby players to a player, the player sends additional requests (attack, movement, skill) to the server for interaction with those nearby players. This increases server load, and the increased load generates more state updates. Consequently, a feedback loop of "transmission$\rightarrow$request$\rightarrow$load$\rightarrow$delay$\rightarrow$re-request" forms, rapidly amplifying the load. More critically, the load contribution differs for each message type. A player death notification is one-time, but nearby player position information triggers continuous tracking requests. Existing load balancing techniques cannot distinguish these type-specific differences or respond to rapid feedback loops that occur within seconds.

\subsection{Related Work}

Research on MMO game server load problems has been conducted in two main environments: (1) DVE/IM research in distributed server environments, (2) efficiency research in single server environments. To understand this research stream, it is necessary to examine the historical context of technological evolution.

Initially, network bandwidth ($100\,\mathrm{Mbps} \sim 1\,\mathrm{Gbps}$) was the main bottleneck, and research naturally developed toward reducing transmitted messages to effectively use limited bandwidth. Distributed Virtual Environment (DVE) and Interest Management (IM) techniques are representative research from this period. Subsequently, as data center networks rapidly advanced to $10 \sim 100\,\mathrm{Gbps}$, network bottlenecks eased, and CPU processing capacity became important due to increased data volume to process. However, with the emergence of cloud computing, horizontal scaling (scale-out) became established as the standard solution, and single server CPU efficiency moved away from academic and industrial interest. Even with this background, \textit{single server CPU processing capacity} still remains as a fundamental scalability limitation\cite{tail-at-scale}.

In this technological evolution process, research naturally divided into two environments: (1) DVE/IM research addressing network bottlenecks between distributed servers, (2) efficiency research addressing CPU processing bottlenecks of single servers. These two environments fundamentally differ in bottleneck location (network vs. CPU), efficiency goals (transmission reduction vs. processing efficiency), and applicable techniques (spatial partitioning vs. parallel processing). Since this research focuses on single server environments, we briefly mention DVE/IM research and focus on single server research, which is the target of this study.

\vspace{1mm}

\textbf{Distributed Server Environment Research (DVE/IM).} Distributed Virtual Environment (DVE) research aims for load distribution among multiple servers and network bandwidth reduction, with Interest Management (IM) techniques being representative\cite{im-survey}. IM reduces network traffic by restricting the information range clients receive through Zone-based (spatial partitioning), Aura-based (subscription area limitation), and Visibility-based (visibility filtering) approaches. However, DVE/IM research is fundamentally different from this study: DVE's main concern is network bottlenecks in distributed environments, while this research aims to control single server CPU load and thread contention. IM techniques only perform message filtering without considering circular causality, cannot adapt to runtime load changes with fixed AOI/policies, and cannot learn the load contribution of each message type.

\vspace{1mm}

\textbf{Load Stabilization Research in Single Server Environments.} Single server research directly related to this study is divided into transmission efficiency and processing efficiency.

(1) \textit{Transmission Efficiency (Smart Reckoning\cite{smart-reckoning})}: Reduces network bandwidth by avoiding message transmission through predicting player movement with distance-based rules and machine learning. It achieved 78.76\% prediction accuracy but cannot dynamically adapt to changing game situations during runtime because it deploys offline-learned models.

(2) \textit{Processing Efficiency (LEARS\cite{lears})}: Improved throughput with a lock-free state model to reduce response time, but when load exceeds CPU core count, thread contention surges sharply (60.36\% in our experiments with Learning OFF), and the fixed structure cannot adapt to runtime load pattern changes.

\vspace{-2mm}
\begin{center}
\begin{minipage}{\textwidth}
\centering
\captionof{table}{Comparison of Existing Research and AFLL}
\label{tab:related-work}
\vspace{-2mm}
\small
\begin{tabular}{@{}llccc@{}}
\toprule
\textbf{Approach} & \textbf{Environment} & \textbf{Goal} & \textbf{Adaptivity} & \textbf{Circular Causality Learning} \\
\midrule
DVE/IM & Distributed & Network messages & Static & X \\
Smart Reckoning & Single & Transmission efficiency & Static & X \\
LEARS & Single & Processing efficiency & Static & X \\
\textbf{AFLL (This work)} & \textbf{Single} & \textbf{CPU/Thread} & \textbf{Dynamic} & \textbf{O} \\
\bottomrule
\end{tabular}
\end{minipage}
\end{center}
\vspace{-2mm}

\subsection{Unresolved Challenges}

The existing research reviewed above pursues different goals but all share the following fundamental limitations: \textbf{(1) Unrecognized Circular Causality} - They do not consider the feedback loop where server-transmitted messages trigger additional client requests, which return as server load. \textbf{(2) Static Policy-Based} - Despite load contribution dynamically changing according to game situations, they only operate with fixed logic (fixed AOI, pre-trained models, lock-free structures) set at development time. \textbf{(3) Absence of Runtime Adaptation Mechanism} - They do not provide mechanisms to dynamically adjust control policies according to load pattern changes during runtime. This is because performing learning algorithms for every transmission decision incurs hundreds of microseconds of delay, which is prohibitive in environments processing tens of thousands of messages per second.

These limitations persist due to several structural factors. First, since the popularization of cloud computing in the 2010s, horizontal scaling became the standard solution, making single server efficiency a topic of low interest. Second, while network improvement is easy to measure and verify, improving CPU load efficiency is difficult to present generalized solutions due to intertwined complex factors, making it less attractive from a publication perspective. Third, circular structures are difficult to capture with single metrics, and existing monitoring tools do not provide mechanisms to learn causal relationships, making the essence of the problem invisible.

In this context, recent research has mainly proceeded in three directions (machine learning-based prediction, auto-scaling, hardware acceleration), all of which are approaches that bypass or indirectly alleviate the CPU bottleneck problem. Consequently, the following two technical challenges remain unsolved: \textbf{(1) Runtime Tracking of Circular Causality} - There is no mechanism to track and control during runtime the feedback loop where server output triggers client input, which returns as server load. \textbf{(2) Implementation of Real-time Adaptation Mechanism} - An architecture that simultaneously achieves ``dynamic adaptation'' and ``real-time performance'' to adapt to load pattern changes in real-time without learning algorithm overhead has not been presented. This research differentiates itself from existing research by solving these two challenges simultaneously.

\subsection{Approach of This Research}

This research proposes Adaptive Feedback Loop Learning (AFLL) to simultaneously solve the two unresolved challenges mentioned above.

\vspace{1mm}

\textbf{Need for Runtime Adaptation.} The load contribution of each message type dynamically changes according to game situations. During large-scale battles, PROJECTILE and DAMAGE surge, but during peaceful exploration, NEARBY\_PLAYERS is the main load cause. Static thresholds cannot respond to these changes: conservative settings excessively block even in peaceful times, degrading game quality, while aggressive settings fail to control during battles, causing performance collapse. Therefore, a mechanism to learn and adapt to load pattern changes during runtime is necessary.

\vspace{1mm}

\textbf{Reason for Choosing Backpropagation.} AFLL uses backpropagation\cite{backprop} to learn the load contribution of each message type. Since server load $L$ is modeled as a linear combination of transmission volume $c_i$ and weight $w_i$ for each message type $i$ ($L = \sum w_i \cdot c_i$), simple linear backpropagation is sufficient without complex deep neural networks. Learned weights directly represent load contribution, making them interpretable, online learning enables runtime adaptation, and stable convergence is guaranteed with momentum and appropriate learning rate.

\vspace{1mm}

\textbf{Overcoming the Real-time Performance Barrier.} Performing backpropagation for every transmission decision incurs hundreds of microseconds of delay, and when processing 20,000 messages per second, 10 seconds of CPU time accumulates, severely degrading performance. This is the technical reason why existing research abandoned runtime learning. AFLL overcomes this barrier by separating learning and control into parallel threads and completely removing learning burden from the message processing path (0ms) through prediction caching (O(k)$\rightarrow$O(1)) and incremental statistics cache (O(n×200)$\rightarrow$O(1)).

\vspace{1mm}

AFLL presents three key differentiators: \textbf{First, Runtime Tracking of Circular Causality} - It dynamically learns the load contribution of each message type while minimizing server load prediction error through backpropagation. It tracks the feedback loop where transmitted messages trigger client requests which return as server load through backpropagation, continuously learning the actual load contribution of each message type during runtime. \textbf{Second, Implementation of Real-time Adaptation Mechanism} - The main message thread immediately decides transmission by only querying learned weights, while the background learning thread separately performs backpropagation. This simultaneously achieves ``dynamic adaptation'' and ``real-time performance''. \textbf{Third, Proactive Load Control} - It presents a proactive load control paradigm that predicts future load with learned load contribution and preemptively suppresses it before reaching the threshold, rather than a reactive approach responding after load occurs. AFLL is designed as a general framework not dependent on specific game server implementations, and in this experiment, we verified its effectiveness by applying it to a LEARS\cite{lears}-based server. Experimental results achieved 48.3\% CPU reduction (6,143ms $\rightarrow$ 3,177ms) and substantial spike reduction (99.7\% reduction in spikes exceeding 50ms) in an environment with 1,000 concurrent users.

\section{AFLL System Design}

\subsection{System Architecture and Operation Flow}

AFLL is an online learning system where real-time learning and control occur simultaneously. The system operates with two parallel threads: (1) the \textit{Main Message Thread} performs a 4-step decision process for each message transmission request to immediately decide transmission, and (2) the \textit{Background Learning Thread} performs backpropagation learning every second to update weights. This enables real-time control without learning overhead (0ms). Figure 1 shows the overall architecture and data flow between components of the AFLL system. The system operates as a circular feedback loop of (1) message transmission $\rightarrow$ (2) load measurement $\rightarrow$ (3) transmission volume recording $\rightarrow$ (4) impact calculation $\rightarrow$ (5) backpropagation learning $\rightarrow$ (6) weight reflection, and the left box summarizes six key features.

\vspace{-2mm}
\begin{center}
\includegraphics[width=0.9\textwidth]{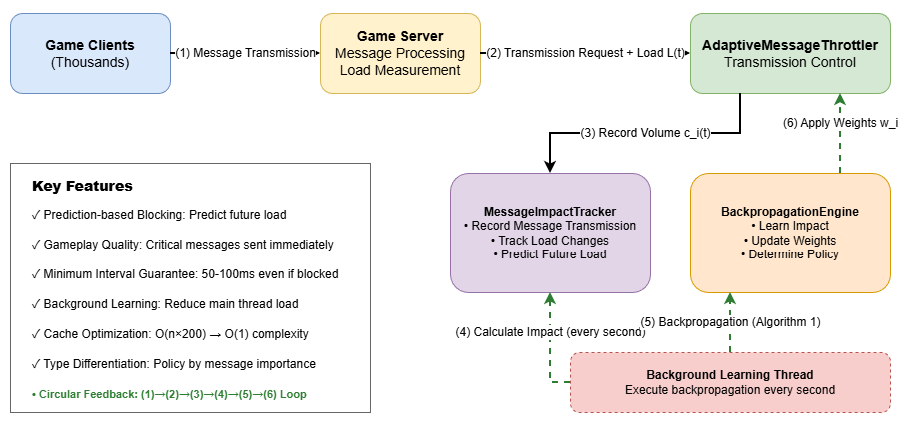}
\captionof{figure}{AFLL System Architecture and Data Flow}
\label{fig:architecture}
\end{center}

\vspace{-1mm}

\textbf{Message Type Definition.} For verification of this research, we implemented an MMO game server prototype and defined 6 message types according to gameplay characteristics (Table 2). AFLL is a general framework applicable to arbitrary message type classifications.

\vspace{-2mm}
\begin{center}
\begin{minipage}{\textwidth}
\centering
\captionof{table}{Message Type Characteristics and Control Policy}
\label{tab:message-types}
\vspace{-2mm}
\small
\begin{tabular}{@{}llccc@{}}
\toprule
\textbf{Type} & \textbf{Description} & \textbf{Interval} & \textbf{Weight} & \textbf{Importance} \\
\midrule
DEATH & Death event & Immediate & [0.05, 0.30] & Critical \\
DAMAGE & Damage amount & Immediate & [0.10, 0.40] & Critical \\
PROJECTILE & Projectile & 50ms & [0.20, 0.70] & Normal \\
CONE & Area attack & 100ms & [0.30, 0.80] & Normal \\
OWN\_STATE & Own position & 100ms & [0.50, 0.95] & Low \\
NEARBY\_PLAYERS & Nearby positions & 200ms & [0.50, 0.95] & Low \\
\bottomrule
\end{tabular}
\end{minipage}
\end{center}
\vspace{-2mm}

DEATH and DAMAGE are essential gameplay information always protected, while the rest are adaptively controlled according to load. Higher weight means greater load contribution and priority blocking during overload.

\vspace{2mm}

\vspace{-2mm}
\begin{center}
\begin{tcolorbox}[colback=gray!5, colframe=black!70, boxrule=0.5pt, arc=2pt, left=3mm, right=3mm, top=2mm, bottom=2mm]
\small
\textbf{Main Notation:}
$w_i$: weight of type $i$,
$c_i(t)$: transmission count of type $i$ at time $t$,
$L(t)$: load score $\in [0,1]$, \\
$\Delta L_{\text{pred}}$: predicted load change,
$\Delta L_{\text{actual}}$: actual load change,
$\alpha=0.03$: learning rate,
$\beta=0.9$: momentum
\end{tcolorbox}
\end{center}
\vspace{-1mm}

\vspace{2mm}

\textbf{Component Roles and Data Flow.} The 4 components in Figure 1 constitute a circular flow with the following roles. (1) \textit{Message Processor} (steps 1-2) receives message transmission requests from game clients, measures queue size, thread contention rate, and memory usage rate to calculate load score $L(t) \in [0,1]$, and passes it to the Throttler (Algorithm 3). (2) \textit{Adaptive Message Throttler} (steps 2-3, 6) combines the weight $w_i$ learned by the Backpropagation Engine with current load $L(t)$ to immediately decide transmission through a 4-step decision process and records transmission events in the Message Impact Tracker (Algorithm 2). (3) \textit{Message Impact Tracker} (steps 3-4) records transmission volume $c_i(t)$ and load change $\Delta L(t)$ for each message type in 100ms windows and efficiently calculates linear regression coefficients with an incremental statistics cache (Algorithm 4). (4) \textit{Backpropagation Engine} (steps 4-6) collects accumulated observation data from the Tracker every second in a background thread, calculates impact, and learns weights $w_i$ that minimize the loss function through backpropagation. Updated weights are reflected in the Throttler and used for the next transmission decision (Algorithm 1). It stably converges using momentum ($\beta=0.9$) and learning rate ($\alpha=0.03$), and type-specific range limits in Table 2 guarantee game quality.

\subsection{Backpropagation Learning Process}

Learning consists of two stages: forward pass and backward pass.

\vspace{1mm}

\textbf{Forward Pass.} At time $t$, predict load change using transmission volume $c_i(t)$ for each message type and learned weight $w_i$:
\begin{equation*}
\Delta L_{\text{pred}}(t) = \sum_{i=1}^{6} w_i \cdot c_i(t)
\end{equation*}

\textbf{Backward Pass.} After 1 second, observe actual load change $\Delta L_{\text{actual}}(t)$ and define prediction error as the loss function. The 1-second delay reflects the feedback loop completion time of circular causality: when the server transmits messages ($\sim$50ms), players receive and process ($\sim$100-200ms), generate and transmit additional requests ($\sim$50ms), and the server receives and processes them ($\sim$100-200ms), the entire cycle completes within an average of about 400-500ms. The 1-second window sufficiently includes this feedback loop. This research only learns output messages that the server transmits to players and does not learn input messages received from players. This is because (1) what the server can directly control is its own output messages, (2) input messages depend on player behavior and cannot be directly controlled by the server, and (3) the main cause of load in MMO games is state broadcasting with $n^2$ complexity. While observed load changes may include other system factors, we assume player reactions to output messages as the main cause of load for learning. The loss function is:
\begin{equation*}
\mathcal{L}(t) = \frac{1}{2}(\Delta L_{\text{actual}}(t) - \Delta L_{\text{pred}}(t))^2
\end{equation*}

Calculate the gradient for each weight $w_i$:
\begin{equation*}
\frac{\partial \mathcal{L}}{\partial w_i} = -(\Delta L_{\text{actual}} - \Delta L_{\text{pred}}) \cdot c_i(t)
\end{equation*}

Update weights with gradient descent including momentum:
\begin{equation*}
\begin{aligned}
v_i(t+1) &= \beta \cdot v_i(t) - \alpha \cdot \frac{\partial \mathcal{L}}{\partial w_i} \\
w_i(t+1) &= w_i(t) + v_i(t+1)
\end{aligned}
\end{equation*}
where $\alpha=0.03$ is the learning rate and $\beta=0.9$ is the momentum coefficient. A learning rate of 0.03 enables fast convergence without excessive overshooting at 1-second learning intervals (convergence observed within about 5 minutes in experiments), and momentum of 0.9 effectively smooths noise from load fluctuations to achieve stable convergence. Algorithm 1 shows the concrete implementation of this backpropagation process.

\vspace{-1mm}
\begin{center}
\small
\textbf{Algorithm 1:} Backpropagation Learning (Background, every 1 second)
\end{center}
\vspace{-3mm}

\begin{tcolorbox}[algobox]   
\begin{AlgorithmCode}
function backpropagation():
  // Forward: prediction calculation
  DeltaL_pred = 0
  for each msgType i in [1..6]:
    c_i = countsSent[i]
    DeltaL_pred += w[i] * c_i

  // Observe actual load change
  DeltaL_actual = L(t) - L(t-1)

  // Backward: gradient and update
  error = DeltaL_actual - DeltaL_pred
  for each msgType i in [1..6]:
    grad = -error * c_i
    velocity[i] = beta*velocity[i] - alpha*grad
    w[i] = w[i] + velocity[i]
    w[i] = clamp(w[i], minW[i], maxW[i])

  // Update prediction cache
  updatePredictionCache(w)
\end{AlgorithmCode}
\end{tcolorbox}
\vspace{-1mm}

Algorithm 1 runs every second in the background thread with the following state management. Whenever the main thread transmits a message, the \textit{Message Impact Tracker} accumulates transmission count $c_i$ by type, and when a 1-second window completes, it passes this accumulated value to Algorithm 1. Load score $L(t)$ is a value normalized to $[0,1]$ range by Algorithm 3 measuring queue size, thread contention rate, and memory usage rate. $L(t-1)$ is stored from $L(t)$ of the previous learning execution, updated as $L(t-1) \leftarrow L(t)$ at the end of each learning cycle. Weight ranges $[\text{minW}_i, \text{maxW}_i]$ are values defined in Table 2, and the clamp operation restricts learned weights from exceeding this range to guarantee game quality. When learning completes, \texttt{updatePredictionCache(w)} is called to update the prediction cache in Algorithm 3, and subsequent transmission decisions of the main thread (Algorithm 2) reference these updated weights.

\vspace{2mm}

\textbf{Learning Example.} At time $t$, if 50 OWN\_STATE messages ($w=0.85$) are transmitted and 1 second later actual load change is observed to be 12.5 larger than predicted, the gradient is $\text{grad} = -12.5 \times 50 = -625$, and through momentum update ($\beta=0.9$, $\alpha=0.03$), $v = 18.75$ and $w = 0.85 + 18.75 = 19.60$, but it is limited to the upper bound (0.95) in Table 2. Thus, weights learn actual load contribution while guaranteeing game quality through range limits.

\subsection{Utilization of Learned Weights}

Weight $w_i$ represents the load contribution of message type $i$. A larger $w_i$ means the message has a greater impact on load and is preferentially controlled during overload. Algorithm 2 shows the 4-step decision process utilizing learned weights.

\vspace{-1mm}
\begin{center}
\small
\textbf{Algorithm 2:} Message Transmission Decision (Main Thread, 0ms overhead)
\end{center}
\vspace{-3mm}

\begin{tcolorbox}[algobox]   
\begin{AlgorithmCode}
function shouldTransmit(msgType, L(t)):
  // Step 0: Protect critical messages
  if msgType in {DEATH, DAMAGE}:
    return TRUE

  // Step 1: Guarantee minimum FPS
  if timeSince >= minInterval[msgType]
     and L(t) < 0.85:
    return TRUE

  // Step 2: Prediction-based allow
  w = getWeight(msgType)
  L_pred = L(t) + w * 1.0
  if L_pred <= 0.70:
    return TRUE

  // Step 3: Probabilistic blocking
  over = (L_pred - 0.70) / 0.30
  P_block = min(w × over × 1.2
                + queuePenalty,
                maxRate[msgType])
  return random() > P_block
\end{AlgorithmCode}
\end{tcolorbox}
\vspace{-1mm}

AFLL uses two overload criteria: \textbf{target load (0.70)} is the safe operating range, \textbf{extreme load (0.85)} is the system threshold. The 4-step decision is as follows.

\textbf{Step 0: Protect Critical Messages.} DEATH and DAMAGE are transmitted immediately regardless of weight to guarantee core gameplay information.

\textbf{Step 1: Guarantee Minimum FPS.} When minimum interval has elapsed and load is below 85\%, force transmission to guarantee minimum transmission period in Table 2.

\textbf{Step 2: Prediction-based Allow.} Allow transmission if predicted load $L_{\text{pred}} = L(t) + w_i$ is below target (0.70).

\textbf{Step 3: Probabilistic Blocking.} If predicted load exceeds target, calculate blocking probability:
\begin{equation*}
P_{\text{block}} = \min\left(w_i \times \frac{L_{\text{pred}} - 0.70}{0.30} \times 1.2 + \text{penalty}_{\text{queue}}, \text{maxRate}_i\right)
\end{equation*}
where $w_i$ is load contribution, 1.2 is a scaling factor, $\text{penalty}_{\text{queue}} = \max(0, (Q-1000)/6666)$ is queue penalty, and $\text{maxRate}_i$ is maximum blocking rate by type. For example, with load 0.80, OWN\_STATE($w=0.85$), and queue 1,200, blocking probability becomes 37\%.

\textbf{Special Handling of NEARBY\_PLAYERS.} Also controls transmission count to dynamically reduce transmission range as load increases. Selects only nearby players by sorting by distance before going through Steps 0-3.

\subsection{Implementation Efficiency}

The Algorithms 1-2 described above present the conceptual design of learning and control, but directly applying them to a real-time game server causes critical performance problems. Performing Algorithm 1's backpropagation learning for every transmission decision incurs hundreds of microseconds of delay, and in environments processing tens of thousands of messages per second, this paralyzes the entire system. As mentioned in the introduction, this overhead is the core reason why existing research abandoned learning-based approaches and relied on static thresholds. This section presents three-stage implementation efficiency techniques to completely eliminate learning overhead.

AFLL is a general learning system not dependent on specific game server implementations. In this experiment, we verified performance improvement effects by applying AFLL to an MMO game server based on the LEARS\cite{lears} lock-free state model. It was implemented in Java 17 and completely eliminated overhead through three stages of implementation efficiency.

\vspace{1mm}

\textbf{Phase 1: Prediction Caching (O(k)$\rightarrow$O(1)).} Instead of recalculating predicted load $\Delta L_{\text{pred}} = \sum w_i c_i$ for every transmission request, recalculate only when weights update (every second) and cache the result. This reduced prediction query time from 125$\mu$s$\rightarrow$8$\mu$s, a 93.6\% reduction.

\vspace{-1mm}
\begin{center}
\small
\textbf{Algorithm 3:} Load Score Calculation and Prediction Caching
\end{center}
\vspace{-3mm}

\begin{tcolorbox}[algobox]   
\begin{AlgorithmCode}
function calculateLoadScore():
  Q = messageQueue.size()
  T = threadContentionRate
  M = memoryUsage / maxMemory
  return 0.6×Q/100 + 0.3×T + 0.1×M

function updatePredictionCache(w):
  // Only called on weight update
  cached = 0
  for each msgType i in [1..6]:
    cached += w[i] * 1.0
  atomicStore(predCache, cached)

function getPredictedLoad(msgType, L(t)):
  // O(1) query for every transmission request
  return L(t) + atomicLoad(predCache)
\end{AlgorithmCode}
\end{tcolorbox}
\vspace{-1mm}

\textbf{Phase 2: Incremental Statistics Cache (Regression O(n×200)$\rightarrow$O(1)).} Linear regression coefficient calculation must iterate through up to 200 windows to calculate $\sum c_i$, $\sum c_i^2$, $\sum c_i \Delta L$ (O(n×200)). The incremental cache immediately updates these sums when windows are added/removed, improving regression coefficient calculation to O(1) time. This reduced statistics calculation time from 2,847$\mu$s$\rightarrow$12$\mu$s, a 99.6\% reduction.

\vspace{-1mm}
\begin{center}
\small
\textbf{Algorithm 4:} Incremental Statistics Cache (O(1) Regression Coefficient Calculation)
\end{center}
\vspace{-3mm}

\begin{tcolorbox}[algobox]   
\begin{AlgorithmCode}
// Global incremental cache
sum_c[i]=0, sum_c2[i]=0
sum_cL[i]=0, windowCount=0

function addWindow(c_i, DeltaL):
  // O(1) update when adding new window
  for each msgType i:
    sum_c[i] += c_i[i]
    sum_c2[i] += c_i[i] * c_i[i]
    sum_cL[i] += c_i[i] * DeltaL
  windowCount++
  if windowCount > 200:
    removeOldestWindow()

function removeOldestWindow():
  // O(1) when removing old window
  old = windows.dequeue()
  for each msgType i:
    sum_c[i] -= old.c_i[i]
    sum_c2[i] -= old.c_i[i]^2
    sum_cL[i] -= old.c_i[i] * old.DeltaL
  windowCount--

function computeRegressionCoeff(i):
  // O(1) regression coefficient (using cached statistics)
  n = windowCount
  num = n*sum_cL[i] - sum_c[i]*sum_DeltaL
  den = n*sum_c2[i] - sum_c[i]^2
  return num / den
\end{AlgorithmCode}
\end{tcolorbox}
\vspace{-1mm}

\textbf{Phase 3: Background Learning (0ms overhead).} The initial implementation had the main thread perform learning immediately after transmission decision, incurring 1,523$\mu$s of learning overhead in the message processing path. Phase 3 moved learning to a completely separate daemon thread. The main thread only records observation data (transmission volume, load) and returns immediately, while the background thread performs backpropagation every second with accumulated data. The two threads communicate through a lock-free queue with no synchronization overhead, and learning executes in parallel on a separate core, completely eliminating learning overhead in the message processing path (0ms).

\begin{center}
\begin{minipage}{\textwidth}
\centering
\captionof{table}{Phase-wise Efficiency Effects}
\label{tab:optimization}
\vspace{-2mm}
\small
\begin{tabular}{@{}lrrr@{}}
\toprule
\textbf{Metric} & \textbf{Initial} & \textbf{Phase 3} & \textbf{Improvement} \\
\midrule
Prediction query & 125 \(\mu\)s & 8 \(\mu\)s & -93.6\% \\
Statistics calculation & 2,847 \(\mu\)s & 12 \(\mu\)s & -99.6\% \\
Learning overhead & 1,523 \(\mu\)s & 0 \(\mu\)s & -100\% \\
\midrule
\textbf{Total} & \textbf{4,495 \(\mu\)s} & \textbf{20 \(\mu\)s} & \textbf{-99.6\%} \\
\bottomrule
\end{tabular}
\end{minipage}
\end{center}

\begin{flushright}
\footnotesize \textit{125$\mu$s, 2,847$\mu$s, 1,523$\mu$s are measured values from Java profiling}
\end{flushright}

\section{Experimental Evaluation}

\textbf{Game Server Implementation.} This research implemented a server prototype simulating a real MMO game environment to verify AFLL's effectiveness. The server was developed in Java 17 and implemented a multi-threaded message processing architecture based on LEARS\cite{lears}'s lock-free state model. The AFLL system was integrated as an independent module on top of this foundation, operating separately as the main message processing thread and background learning thread. Simulation clients were implemented with a probability model to reproduce actual player behavior, with each simulation client executing in an independent thread to reproduce the concurrency of real network environments.

\vspace{1mm}

\textbf{Experimental Environment and Settings.} Experiments were conducted on Intel Core i7-12700K (12 cores), 32GB RAM, Windows 11 Pro environment with 1,000 concurrent users, each for 30 minutes. Each client independently generates movement input (30\% probability for each direction), Fire input (20\% probability), and Cone input (10\% probability) for every transmission, reproducing behavior patterns where players continuously move and intermittently attack in actual MMO games. Load was generated with a 60Hz burst pattern after 30 seconds, with target server load set to 0.70. We used learning rate ($\alpha$) of 0.03, momentum ($\beta$) of 0.9, and learning interval of 1 second. Each message type starts with differentiated initial weights according to importance and is learned according to actual load contribution through backpropagation.

\vspace{1mm}

\textbf{Metric Measurement Method.} Thread Contention Rate is defined as the ratio of active thread count to maximum thread count: $T = \text{activeThreads} / \text{maxThreads}$. The closer the value is to 1.0 (100\%), the more all threads are simultaneously active, meaning context switching and synchronization overhead surge. Load score $L(t)$ is calculated as defined in Algorithm 3.

\begin{center}
\centering
\includegraphics[width=0.9\textwidth]{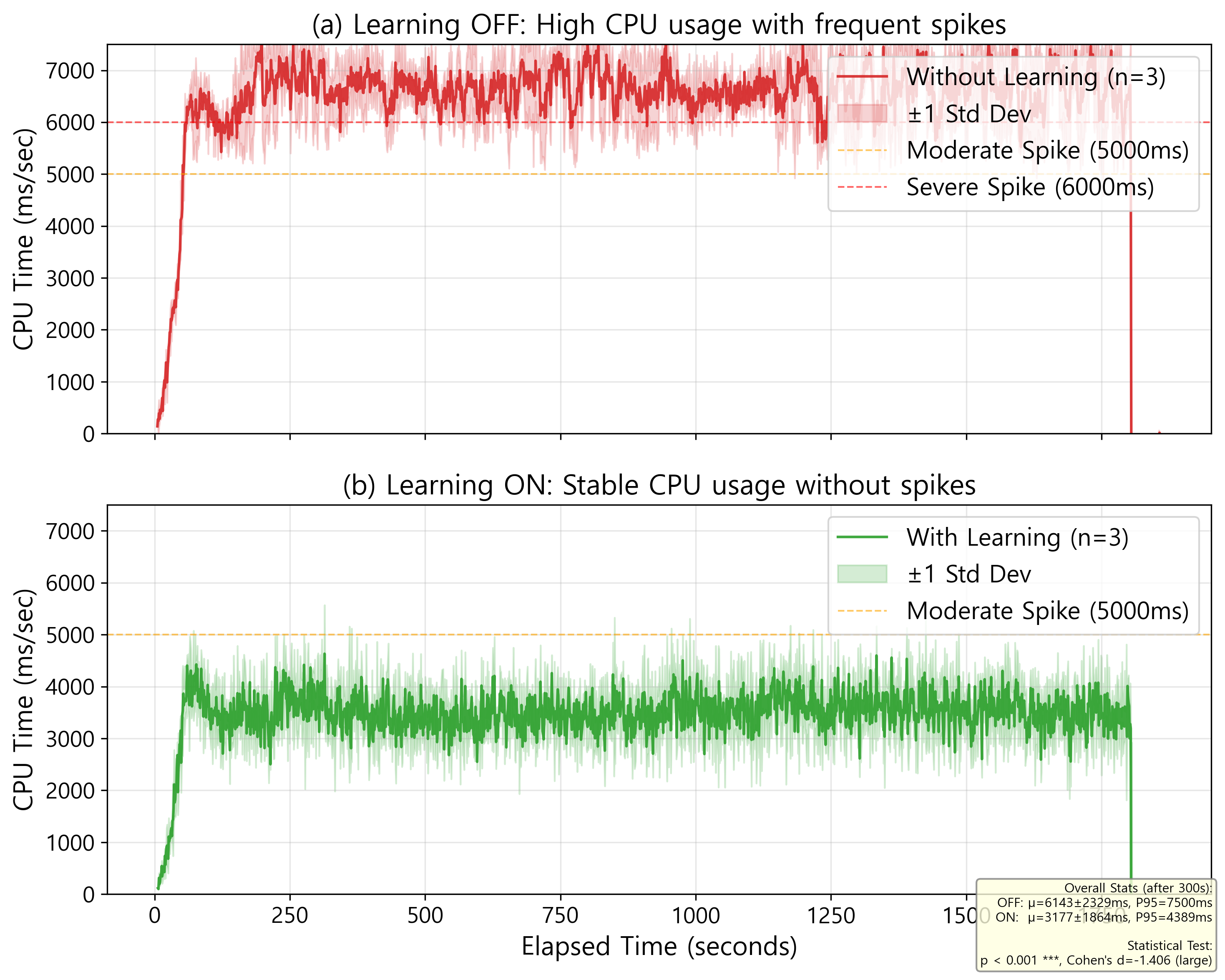}
\captionof{figure}{CPU Time Series Comparison (30 minutes, n=3 mean±std)}
\label{fig:cpu-timeseries}
\end{center}

Figure 2 shows CPU time changes over 30 minutes. For Learning OFF, large performance degradations occur frequently, while Learning ON maintains a stable pattern. According to the statistical analysis results in Table 4, Learning OFF averaged 6,143ms/sec (P95 7,500ms, P99 7,706ms), while Learning ON achieved an average of 3,177ms/sec (P95 4,389ms, P99 4,758ms) ($p < 0.001$, Cohen's $d = -1.41$, large effect). Particularly noteworthy is that while frequent performance spikes (CPU time over 5,000ms) occurred throughout the entire experiment duration in Learning OFF, Learning ON maintains stable performance. Learning OFF had 94.5\% (5,250 cases) of spikes over 5,000ms and 30.7\% (1,706 cases) over 7,000ms, but Learning ON had only 0.3\% (14 cases) over 5,000ms and completely eliminated spikes over 6,000ms.

\vspace{-3mm}
\begin{center}
\begin{minipage}{\textwidth}
\centering
\captionof{table}{CPU Performance and Thread Contention Statistics Comparison}
\label{tab:performance}
\vspace{-2mm}
\small
\begin{tabular}{@{}lrrr@{}}
\toprule
\textbf{Metric} & \textbf{Learning OFF*} & \textbf{Learning ON*} & \textbf{Improvement} \\
\midrule
\multicolumn{4}{c}{\textit{CPU Performance (ms/sec)}} \\
Mean & 6,143 & 3,177 & \textbf{-48.3\%} \\
Median & 6,625 & 3,468 & -47.7\% \\
Std. Dev. & 2,329 & 1,864 & -20.0\% \\
P95 & 7,500 & 4,389 & -41.5\% \\
P99 & 7,706 & 4,758 & -38.2\% \\
\midrule
\multicolumn{4}{c}{\textit{Thread Contention (\%)}} \\
Mean & 60.36\% & 21.51\% & \textbf{-64.4\%} \\
Median & 66.70\% & 4.20\% & -93.7\% \\
100\% occurrence & 12.49\% & 0.40\% & -96.8\% \\
\midrule
\multicolumn{4}{l}{\footnotesize * Both Learning OFF/ON are average values from three independent experiments (see Table 5 for reproducibility statistics)} \\
\bottomrule
\end{tabular}
\end{minipage}
\end{center}

Table 4 shows CPU performance statistics. Mean CPU time decreased by 48.3\%, and median decreased by 47.7\%, showing consistent performance improvement. P95 and P99 decreased by 41.5\% and 38.2\% respectively. Standard deviation decreased from 2,329ms to 1,864ms, a 20.0\% reduction, improving performance stability. In terms of thread contention, Learning OFF showed a high average contention rate of 60.36\%, while Learning ON decreased to 21.51\%, a 64.4\% reduction. Particularly, the 100\% contention state occurrence rate decreased from 12.49\% to 0.40\%, a 96.8\% reduction, nearly eliminating extreme contention situations.

\begin{center}
\centering
\includegraphics[width=0.9\textwidth]{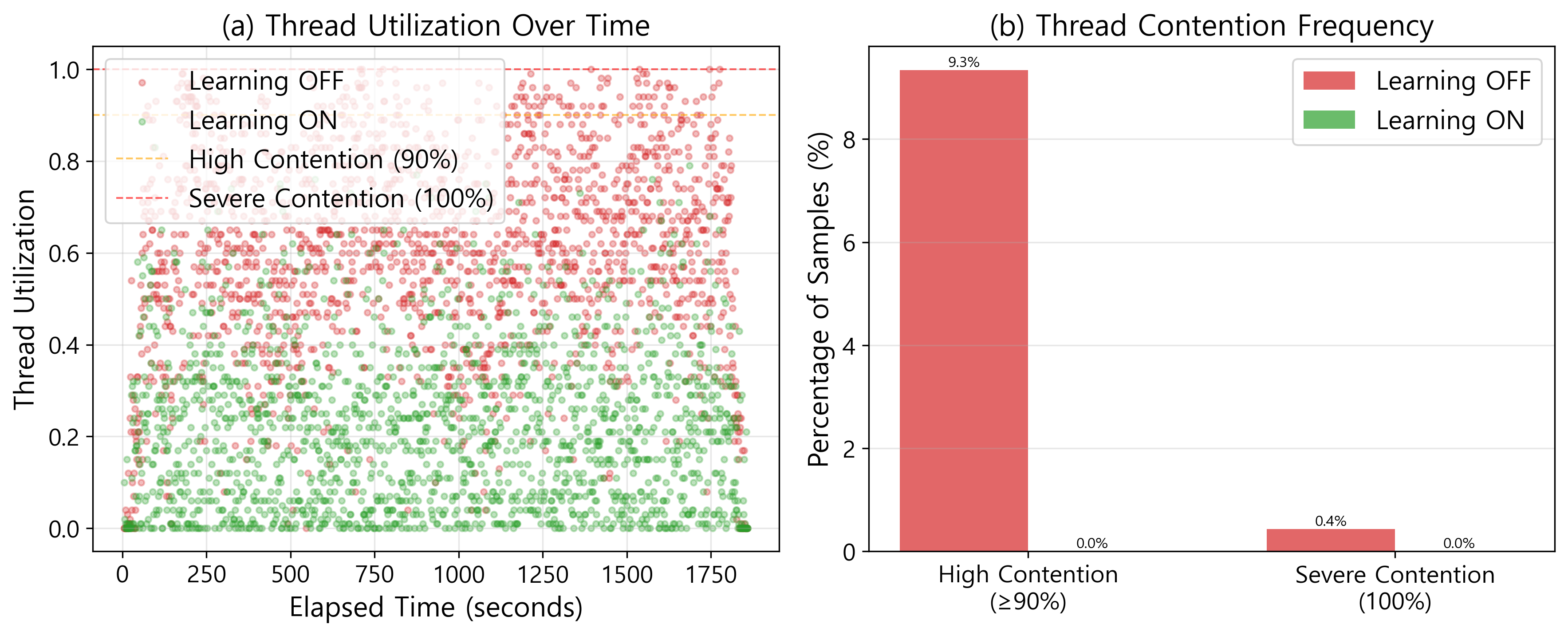}
\captionof{figure}{Thread Contention Analysis and Distribution}
\label{fig:contention}
\end{center}

Figure 3 shows the time series change and distribution of thread contention. Learning OFF has considerable samples distributed in the high contention range (80-100\%), with 100\% contention state occurring about 694 times (12.49\%). In contrast, Learning ON has most samples (98.7\%) concentrated in the low contention range (0-20\%), with 100\% contention reduced to about 22 times (0.40\%).

\begin{center}
\centering
\includegraphics[width=0.9\textwidth]{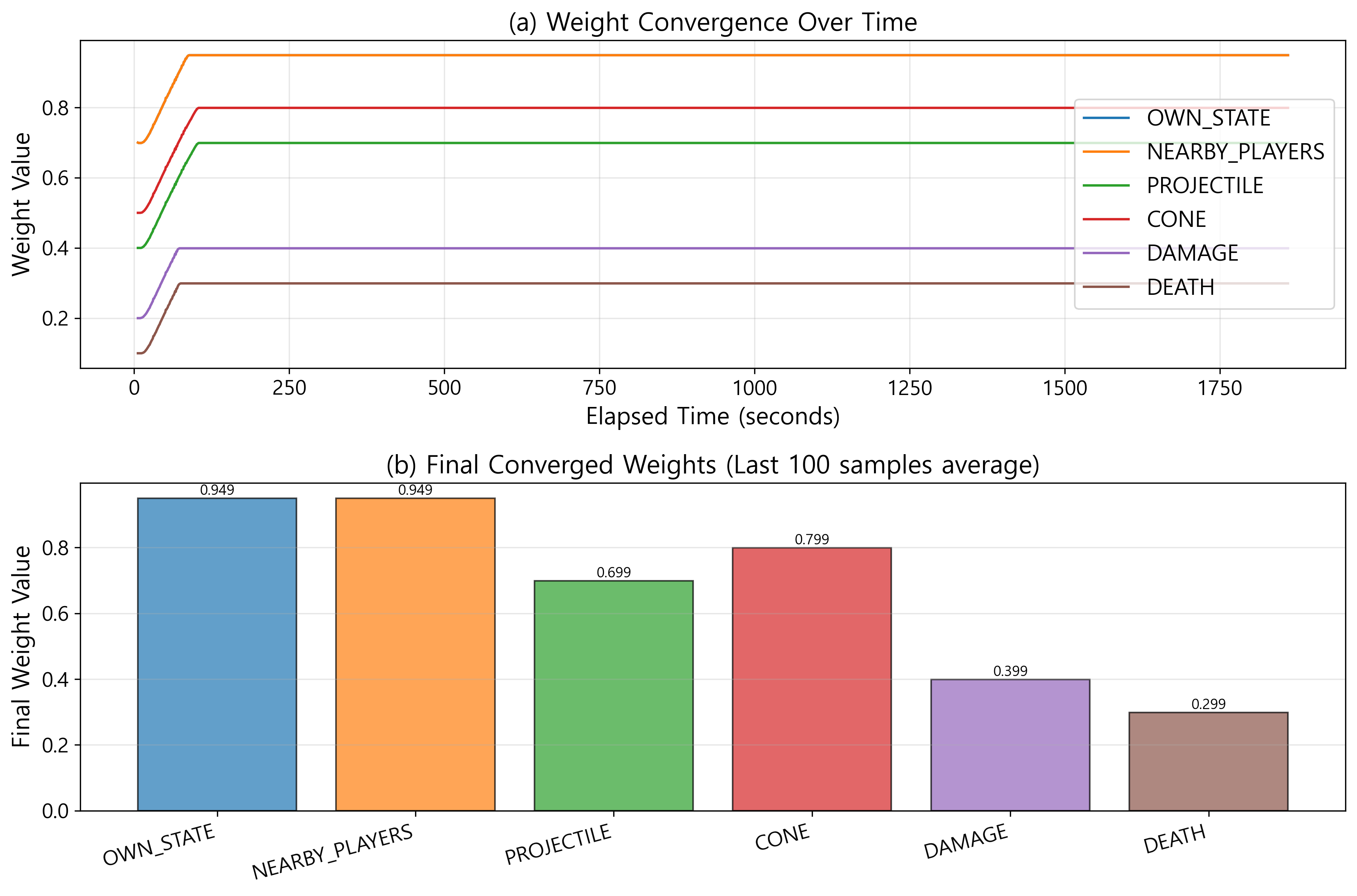}
\captionof{figure}{Weight Convergence Process by Message Type}
\label{fig:weights}
\end{center}

Figure 4 shows weight changes for each message type during the learning process. All weights converged to stable values within about 5 minutes. OWN\_STATE (0.949) and NEARBY\_PLAYERS (0.949) showed the highest final weights, followed in decreasing order by CONE (0.799), PROJECTILE (0.699), DAMAGE (0.399), and DEATH (0.299). Weights reflect the impact of each message on load, with messages having high weights preferentially controlled during load spikes. Initial weights were differentiated according to message importance and adapted to actual load patterns through learning.

Performance spikes (CPU time over 5,000ms within 1-second intervals) occurred 5,250 times (94.5\%) in Learning OFF but decreased to 14 times (0.3\%) in Learning ON. Extreme spikes over 6,000ms were completely eliminated from 4,282 times (77.0\%) in Learning OFF to 0 times in Learning ON.

\vspace{1mm}

\textbf{Reproducibility and Statistical Validation.} Both Learning OFF and Learning ON performed three independent experiments under identical conditions. Learning OFF operated with fixed weights, showing stable reproducibility with a coefficient of variation (CV) of 1.83\%. Interestingly, Learning ON showed lower CV despite probabilistic elements, with all three experiments converging to identical weights (0.949, 0.949, 0.799, 0.699, 0.399, 0.299), and Table 5 shows extremely stable consistency with CPU time CV of 0.54\%, CPU P95 of 0.44\% (both under 1\%), and thread contention of 1.43\% (under 2\%). This demonstrates that the baseline is very stable and AFLL stably converges regardless of initial conditions or system variations. The performance difference between the two conditions is statistically very significant with independent sample t-test result $p < 0.001$, and Cohen's $d = -1.41$ (large effect) shows very large improvement from a practical perspective.

\begin{center}
\begin{minipage}{\textwidth}
\centering
\captionof{table}{Comparison of Three Independent Experiments and Reproducibility Analysis}
\label{tab:reproducibility}
\vspace{-2mm}
\small
\begin{tabular}{@{}lcccccc@{}}
\toprule
\textbf{Metric} & \textbf{Exp1} & \textbf{Exp2} & \textbf{Exp3} & \textbf{Mean$\pm$SD} & \textbf{Range} & \textbf{CV} \\
\midrule
CPU Mean & 3,155 & 3,181 & 3,196 & 3,177$\pm$17 & 41 & 0.54 \\
CPU P95 & 4,362 & 4,409 & 4,382 & 4,384$\pm$19 & 47 & 0.44 \\
Thread Contention & 21.81 & 21.08 & 21.62 & 21.51$\pm$0.31 & 0.73 & 1.43 \\
Blocking Rate & 59.97 & 60.11 & 60.16 & 60.08$\pm$0.10 & 0.19 & 0.16 \\
\midrule
\multicolumn{7}{@{}l@{}}{\scriptsize * CPU unit: ms, Thread contention/Blocking rate: \%, Range: max-min, CV: coefficient of variation(\%)} \\
\multicolumn{7}{@{}l@{}}{\scriptsize * Weights: OWN\_STATE(0.949), NEARBY\_PLAYERS(0.949), CONE(0.799),} \\
\multicolumn{7}{@{}l@{}}{\scriptsize \hspace{2.5mm} PROJECTILE(0.699), DAMAGE(0.399), DEATH(0.299) - Completely identical in all three experiments} \\
\bottomrule
\end{tabular}
\end{minipage}
\end{center}

\section{Causal Relationship Analysis}

AFLL's performance improvement is explained through a 3-stage causal chain. \textbf{Stage 1 (Identical Input)}: Game activities in both conditions are essentially identical. Player count (972.6 vs 973.4, 0.1\% difference), visibility range (727.2 vs 726.7, 0.1\% difference), and game events (around 1\% difference) are all similar, so the input load the server must process is virtually identical. \textbf{Stage 2 (Output Control 81.5\% Reduction)}: Learning OFF transmits all state changes (17,444 msg/sec), while Learning ON reduces by 81.5\% (3,227 msg/sec) through selective transmission. The key is \textit{protecting game-essential information while controlling only load amplification factors}. Damage and death events guarantee 100\% transmission to maintain game fairness, while only nearby player position information is selectively reduced (100\%$\rightarrow$24\%, distance-based priority). Although AFLL blocks 60.1\% of message attempts, the key is not the blocking itself but \textit{the effect of breaking the circular feedback loop}. As blocked messages reduce information clients receive, subsequent client responses also decrease, which further lowers server load, creating a virtuous cycle. Consequently, it achieves an amplification effect of reducing total message transmission by 81.5\% compared to Learning OFF. \textbf{Stage 3 (CPU 48.3\% Reduction)}: Message reduction reduces load through direct effects (transmission processing cost) and indirect effects (concurrency overhead), with indirect effects being larger. Thread contention reduction of 64.4\% (60.36\%$\rightarrow$21.51\%) causes non-linear overhead to plummet.

\begin{center}
\centering
\includegraphics[width=0.9\textwidth]{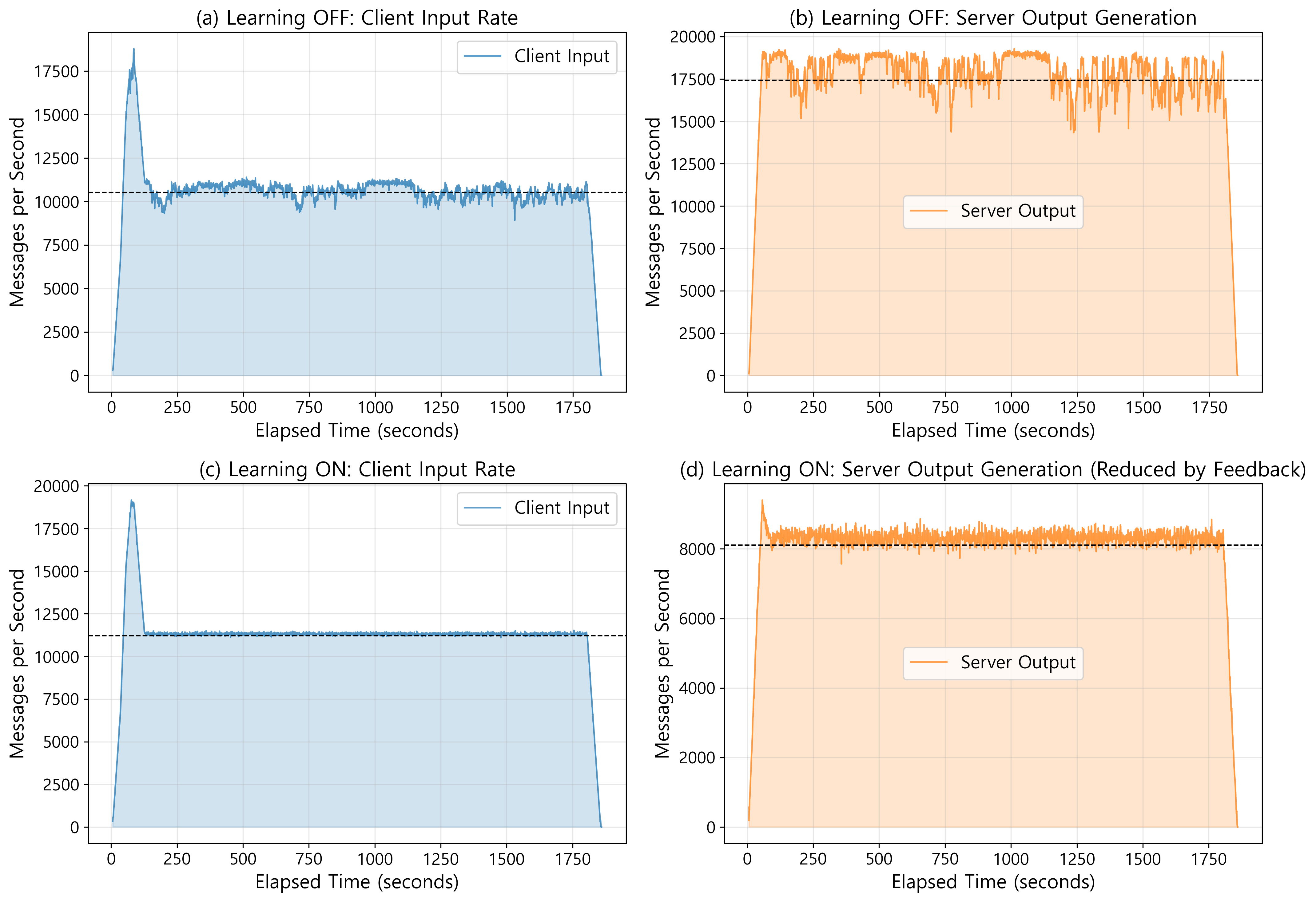}
\captionof{figure}{Input-Output Flow and 3-Stage Causal Chain}
\label{fig:io-flow}
\end{center}

Figure 5 shows the complete causal chain. Input (player count, game activity) shows identical patterns in both conditions. The spike where client input messages increase to 1.8 times the average during the first 5 minutes reflects active activity in the early connection phase and appears identically in both Learning OFF and ON. However, AFLL intelligently controls output (message transmission) for identical input to reduce load.

\begin{center}
\centering
\includegraphics[width=0.9\textwidth]{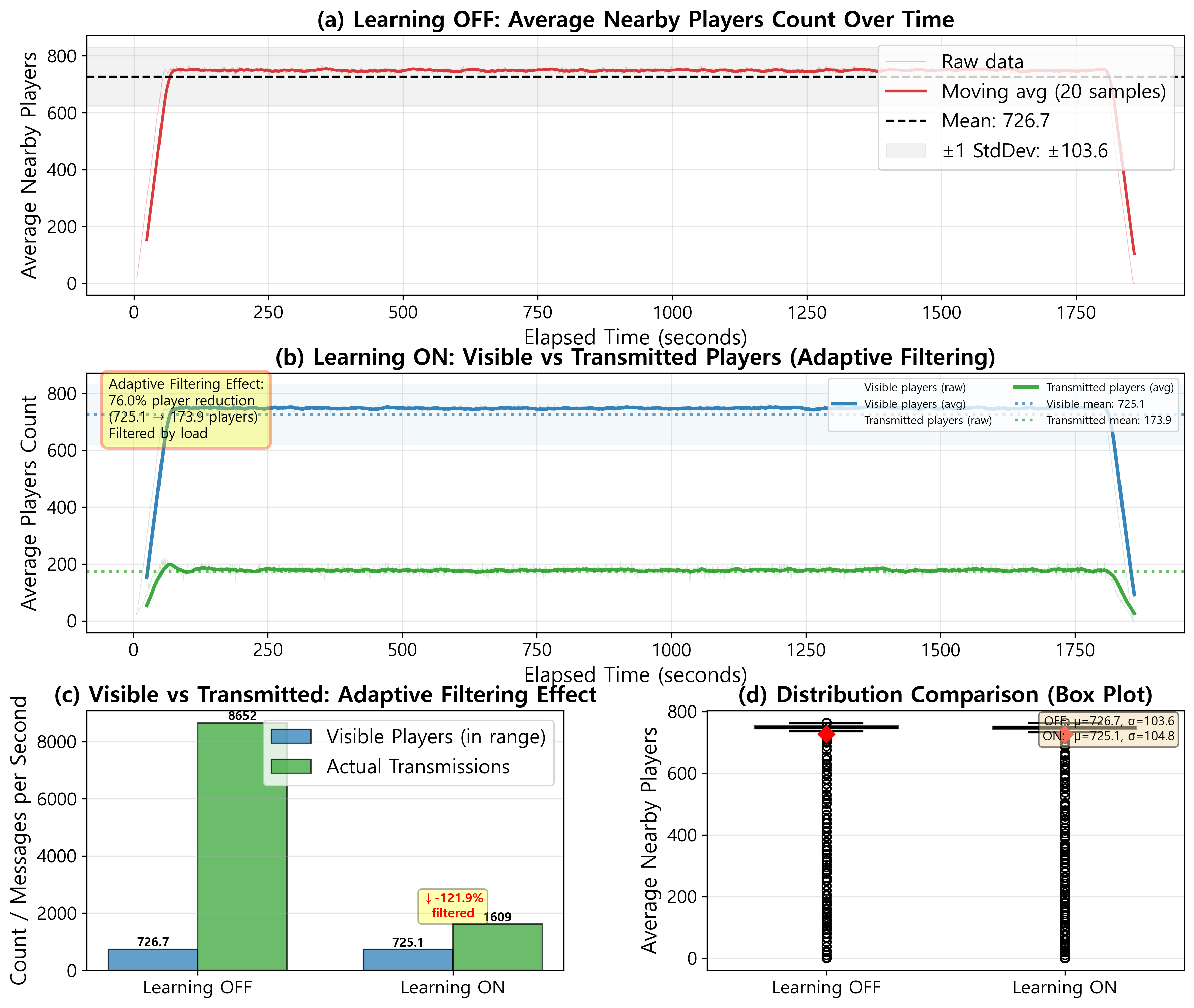}
\captionof{figure}{Nearby Player Visibility vs Transmission Rate Comparison}
\label{fig:nearby-players}
\end{center}

Figure 6 shows the core mechanism of output control. Learning OFF transmits 100\% of information about all nearby players (727.2 people) within visibility range. In contrast, Learning ON selects with distance-based priority from the same visibility range (726.7 people) to transmit only an average of 174.3 people (24\%). Nearby players are guaranteed with high priority, while distant players are omitted when load is high. This reduces load without degrading game experience by maintaining near-range information that players actually interact with while only controlling distant NEARBY\_PLAYERS messages, which are the core link of circular feedback. Despite adding learning overhead (38ms/sec, 1.3\%), it achieves net profit (2,928ms, 48.3\%) through concurrency overhead savings (2,966ms).

\begin{center}
\centering
\includegraphics[width=0.9\textwidth]{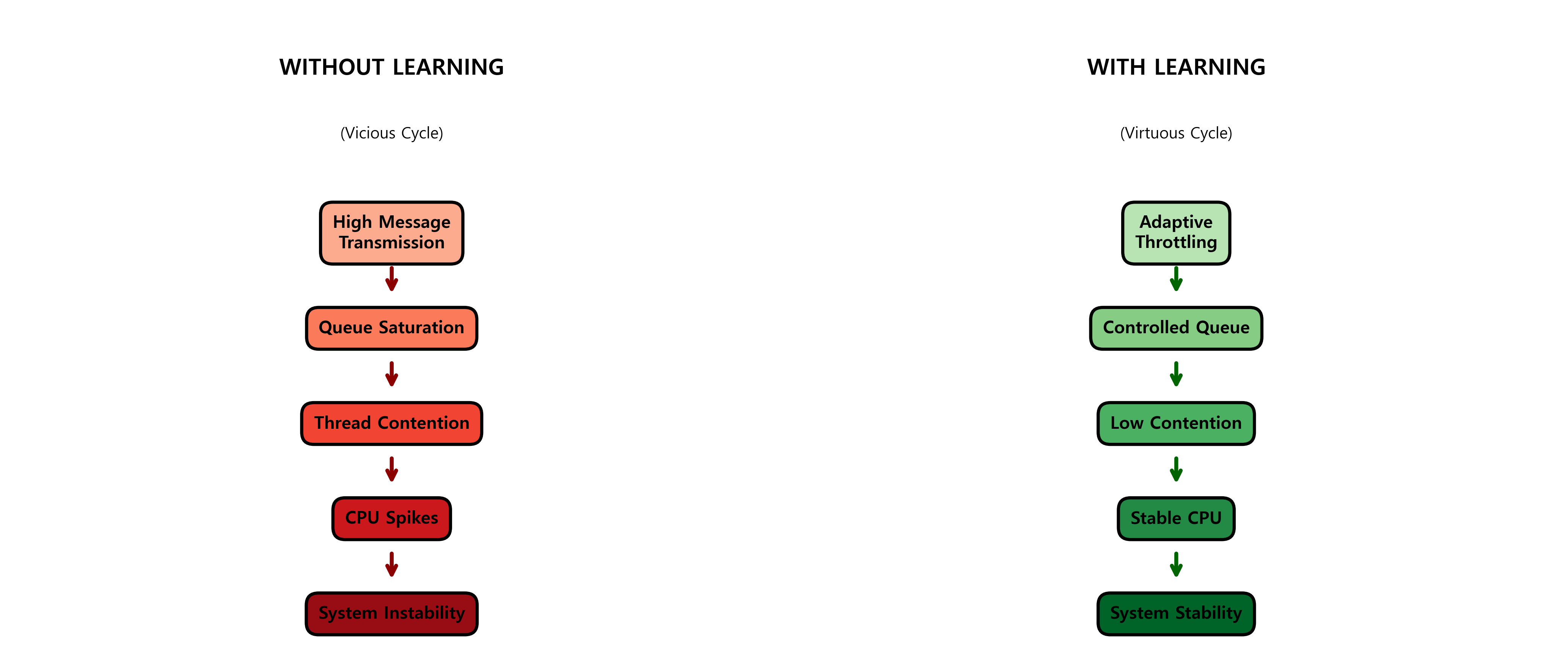}
\captionof{figure}{Load Control Mechanism: Non-linear Overhead}
\label{fig:mechanism}
\end{center}

Figure 7 shows that CPU time surges non-linearly when exceeding the load threshold. Below the observed threshold (about 70\% load), CPU time is roughly linearly proportional to load, but above the threshold, thread contention sharply increases and CPU time surges non-linearly. Learning OFF operates in the non-linear region with load exceeding 100\%, where concurrency overhead exceeds actual work cost. AFLL guarantees operation in the linear region by keeping load below the threshold (70\%).

This phenomenon can be explained by the traffic light paradox. At a busy intersection without traffic lights, if all vehicles try to enter simultaneously, deadlock occurs and average transit time increases. Conversely, when entry is controlled with traffic lights, individual vehicles wait but overall transit time actually decreases. Similarly in game servers, AFLL improves overall system performance by appropriately controlling message transmission to prevent thread contention.

\begin{center}
\centering
\includegraphics[width=0.9\textwidth]{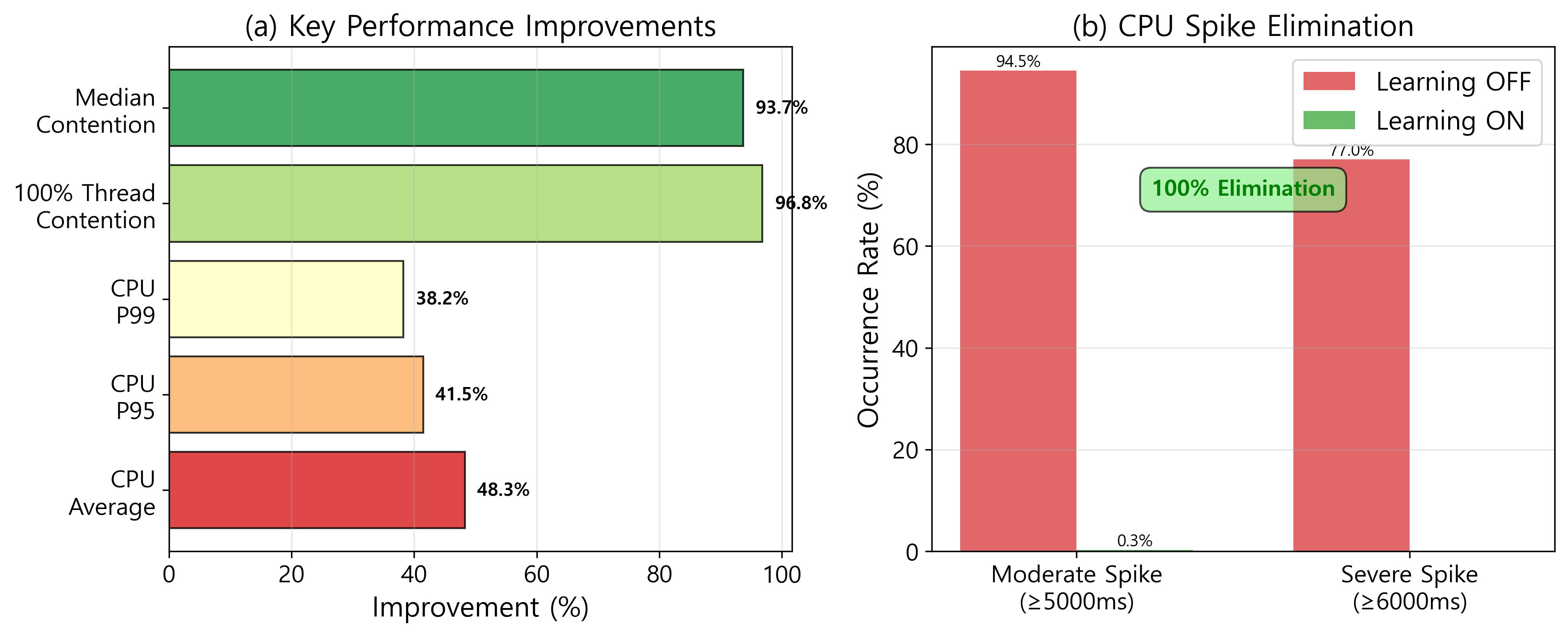}
\captionof{figure}{Comprehensive Comparison of Main Metrics}
\label{fig:summary}
\end{center}

Figure 8 shows a comprehensive comparison of main metrics. AFLL showed dramatic improvement in all major metrics including CPU time, thread contention, and performance spikes. Particularly, learning overhead (38ms/sec) is only 1.3\% of total CPU savings (2,966ms/sec), confirming very high return on investment.

\section{Conclusion}

This paper proposed the AFLL system that learns and controls circular causality in MMO game servers. The main contributions of this research are as follows. First, to the best of our knowledge, we proposed the first real-time adaptive system that learns circular causality in MMO game servers to proactively control load. While existing research relied on static threshold-based throttling or post-processing efficiency, AFLL learns the load contribution of each message type through backpropagation to predict and preemptively block load spikes. Second, we presented implementation efficiency techniques that achieve 0ms overhead by parallelizing learning and control. We completely removed learning burden from the message processing path through prediction caching (O(k)$\rightarrow$O(1)), incremental statistics cache (regression calculation O(n×200)$\rightarrow$O(1)), and background learning separation. Third, we demonstrated 48.3\% CPU reduction (6,143ms $\rightarrow$ 3,177ms), 64.4\% thread contention reduction (60.36\% $\rightarrow$ 21.51\%), and near-complete elimination of performance spikes in an environment with 1,000 concurrent users. Fourth, we verified excellent reproducibility for both Learning OFF and Learning ON (coefficients of variation 1.83\% and 0.54\%), with three Learning ON experiments converging to identical weights. Fifth, we identified achieving net profit (2,928ms) through concurrency overhead savings (2,966ms) while adding learning overhead (38ms/sec, 1.3\%). These results show that AFLL can be applied to production environments.

\vspace{2mm}

\textbf{Research Scope and Future Challenges.} This research focused on securing load stability and selected message transmission load that rapidly amplifies through circular causality as the learning target. Other internal processing load factors such as unpredictable or periodically occurring game logic and AI computation were excluded from learning targets. This is because AFLL's goal is not overall load reduction but prevention of rapid load amplification due to feedback loops.

Additional considerations exist in actual production environments. First, if the proportion of non-message load factors is larger, we can (1) adjust learning window size to mitigate interference from other system loads, or (2) separate system load into message-related and non-message-related loads for learning (e.g., separately measure CPU time of message processing threads and game logic threads). Second, CPU load may be measured lower than actual due to external environmental factors such as thread lock contention, I/O waiting, and external system failures. In such cases, the increase rate of work queue ($\Delta Q / \Delta t$) can be utilized as an auxiliary weight in load score $L(t)$ to improve learning accuracy.

Future research plans to explore the following: (1) \textbf{Multi-dimensional Load Metrics} - Learning load vector $\mathbf{L}(t) = [L_{\text{cpu}}, L_{\text{net}}, L_{\text{mem}}, L_{\text{disk}}]$ for independent control of CPU, network, memory, and disk. For example, selectively block only network-intensive messages when CPU is relaxed but network is saturated. (2) \textbf{Adaptive Learning} - Dynamically adjust learning rate and window size according to load pattern volatility to simultaneously achieve fast convergence and stability. (3) \textbf{Deep Neural Network Extension} - Consider applying time series models such as LSTM and Transformer for learning complex non-linear relationships. However, only consider when linear models are insufficient due to high learning cost and difficulty in interpretation. (4) \textbf{Distributed Environment Extension} - Share weights between servers in multi-server environments to achieve global optimization. Learn independently if there are local characteristics, and aggregate weights if there are common patterns.

\vspace{2mm}

\textbf{Scalability and Domain-specific Considerations.} AFLL's core principle of circular causality learning is applicable beyond MMO game servers to other real-time systems. It can be utilized in systems with feedback loops such as social networks (notification$\rightarrow$user reaction$\rightarrow$additional notification), IoT (sensor update$\rightarrow$control command$\rightarrow$sensor remeasurement), financial trading (market data$\rightarrow$trading order$\rightarrow$market change), and cloud auto-scaling (resource expansion$\rightarrow$load migration$\rightarrow$additional expansion).

However, unlike MMO games, message filtering may not be safe in other domains. In games, even if some messages (NEARBY\_PLAYERS) are blocked, there is little impact on gameplay, but in financial trading (order confirmation), IoT control (emergency alerts), and medical monitoring (vital signs), message loss is fatal to integrity and safety. Therefore, domain application requires the following: (1) \textbf{Fine-grained Priority System} - Subdivide message importance by domain (e.g., in financial systems: transaction confirmation $>$ market data $>$ statistical information) to block only low priority. (2) \textbf{Alternative Control Mechanisms} - Instead of message blocking, apply domain-appropriate methods such as processing delay (request batching), sampling frequency adjustment, and resource allocation adjustment. For example, in IoT, dynamically adjust sensor update cycles. (3) \textbf{Safety Verification} - Verify in real-time that learned control policies do not compromise system safety. For example, in medical systems, trigger alerts and disable control if critical messages are delayed beyond a certain time. Through such enhancements, AFLL's circular causality learning principle is expected to contribute to load stabilization of various real-time systems.

\end{document}